\documentclass[sigconf, nonacm]{acmart}

\usepackage{pvldb}

\renewcommand\vldbdoi{XX.XX/XXX.XX}
\renewcommand\vldbpages{XXX-XXX}
\renewcommand\vldbavailabilityurl{}

\usepackage{balance}

\usepackage{booktabs}
\usepackage{xspace}
\usepackage{amsmath}
\usepackage{graphicx}
\usepackage{tikz}
\usetikzlibrary{arrows.meta,positioning,fit,backgrounds}
\usepackage[ruled,noend,linesnumbered]{algorithm2e}

\newcommand{\G}{\ensuremath{G}\xspace}
\newcommand{\A}{\ensuremath{\mathcal{A}}\xspace}
\newcommand{\R}{\ensuremath{\mathcal{R}}\xspace}
\newcommand{\Hs}{\ensuremath{\mathcal{H}}\xspace}
\newcommand{\Sset}{\ensuremath{S}\xspace}
\newcommand{\cost}{\ensuremath{\mathit{cost}}}
\newcommand{\an}{\ensuremath{\alpha}}
\newcommand{\ctoc}{c-to-c\xspace}
\newcommand{\ctov}{c-to-v\xspace}
\newcommand{\vtoc}{v-to-c\xspace}
\newcommand{\vtov}{v-to-v\xspace}
\newif\ifdraft
\drafttrue
\ifdraft
  \newcommand{\todo}[1]{\textcolor{red}{[TODO: #1]}}
  \newcommand{\claudecolor}{\color{violet}}       
  \newcommand{\gonzalo}[1]{\textcolor{green!50!black}{Gonzalo says: #1}}
\else
  \newcommand{\todo}[1]{}

  \newcommand{\claudecolor}{}
  \newcommand{\gonzalo}[1]{}
\fi

\begin{document}

\title[Tunnels: Compressing the Solution Subgraph of Regular Path Queries]{Tunnels: Compressing the Solution Subgraph \\ of Regular Path Queries}

\author{Adri\'an G\'omez-Brand\'on}
\affiliation{\institution{Universidade da Coruña \& IMFD}\city{A Coruña}\country{Spain}}
\email{adrian.gbrandon@udc.es}
\author{Gonzalo Navarro}
\affiliation{\institution{DCC, University of Chile \& IMFD}\city{Santiago}\country{Chile}}
\email{gnavarro@uchile.cl}

\begin{abstract}
Regular path queries (RPQs) are increasingly expected to return the paths that match them, not
only their endpoints. Since the number of matching paths can be exponential or infinite, a
standard way to deliver them (PMR) is the subgraph $\Hs$ of the product between the graph and the
query automaton that contains exactly the nodes lying on some solution path. We show that $\Hs$
can be compressed much further when the graph is available to the decoder: most of
$\Hs$ can be regenerated by re-running small parts of the traversal, and only the places where the traversal
would waste significant effort on dead branches need to be stored. We formalize this as \emph{tunnels}:
regions of $\Hs$ that are re-expanded from a stored entry point with a bounded amount of extra
work, either an absolute budget or one proportional to the output produced. We handle the three
classes of RPQs: constant-to-variable, constant-to-constant, and variable-to-variable. In the latter, we study the combination of tunnels with cutting
the query at a selective articulation point of the automaton and encoding the answer as a union of Cartesian products of leftward and rightward traversals. Our experimental results show improvements of 2 to 3 orders of magnitude over the baseline space of PMR, for an acceptable increase in traversal time.
\end{abstract}

\maketitle

\vldbtopmatter

\section{Introduction}

Regular path queries (RPQs) select pairs of nodes of an edge-labeled graph connected by a path
whose label sequence matches a regular expression. They are the navigational core of SPARQL
property paths and of the path patterns of GQL and SQL/PGQ~\cite{DeutschEtAl22}, and they are
heavily used in practice: about 24\% of the queries in the Wikidata logs use property
paths~\cite{BMT19}, a recent study found 148.7 million RPQs among 937.2 million logged SPARQL
queries~\cite{HM25}, and transitive navigation such as \texttt{P31/P279*} (instance of a subclass
of) is among the most common patterns~\cite{BMT19,MalyshevEtAl18,HM25}.

The new standards have moved the focus from returning pairs of endpoints to returning the
\emph{paths} themselves: GQL and SQL/PGQ let path patterns bind and return the matched
paths~\cite{DeutschEtAl22}, and recent engines return them for all the solutions of a
query~\cite{MNPRVV23,FMRV24}. This is challenging because the number of matching paths between two
nodes can be exponential in the path length, or even infinite when cycles are allowed.

A natural compact representation of the output is the \emph{solution subgraph}. Evaluating an RPQ amounts to a traversal
of the product $\G\times\A$ of the graph and an automaton for the expression. Keeping only the
product nodes that lie on some path from the source to an accepting node yields a subgraph $\Hs$
in which \emph{every} path from the root to an accepting node is a solution, and every solution
is such a path. This is the idea behind the \emph{path multiset representations} (PMRs) of
Martens et al.~\cite{MNPRVV23} (see also~\cite{KMN23}): a graph whose nodes are mapped homomorphically to the nodes of the
database, with start and target nodes, so that its source-to-target paths spell the matching
paths. PMRs represent exponentially or infinitely many
paths in compacted form, the paths can be enumerated from them, and
experimental studies show that they drastically speed up the evaluation of path queries in
graph database engines~\cite{MNPRVV23,FMRV24}. The subgraph $\Hs$ is the PMR that the
product-graph evaluation of an RPQ produces naturally.

In large knowledge graphs, however, $\Hs$ itself is large. A query like \texttt{?x P31/P279* C}
evaluated from a popular class $C$ reaches millions of instances. On our Wikidata graph, the
plain PMR of a logged variable-to-variable query takes 158 MB on average and up to 1.7 GB
(that of a constant-to-variable query takes up to 13 MB); a system that serves or caches
the results of many queries at once multiplies these figures.
Our main observation is that most of that structure is redundant with the graph itself: if whoever
consumes the paths has the graph at hand, they can traverse it again. A PMR
can then be represented much more succinctly by trading size for partial retraversing. How much traversing costs depends on how the
graph is represented; we measure it on various compact representations~\cite{AHNRRS21,AHNR22,NR24} and on a plain one.
The idea parallels the compressed representations of conjunctive query results of
Deep and Koutris~\cite{DK18}, which leave parts of the result implicit when recomputing them
from the database costs about as much as traversing them, thereby trading space for answer time. 

In an extreme, fully re-running the traversal regenerates $\Hs$; what makes re-running expensive
is not $\Hs$ but the \emph{dead} part of the traversal, the branches that are explored but do not
lead to any solution. In many queries those dead branches are rare, or short, or concentrated
in a few nodes. Storing information only at those nodes, and letting the decoder regenerate the
rest at a cost close to that of traversing $\Hs$, can shrink the representation by orders of magnitude while keeping the extra work bounded.

\paragraph{Contributions.}
\begin{itemize}
\item We show that encoding $\Hs$ relative to the graph, identifying each successor by its position among the neighbors of its parent and gap-encoding the lists, reduces it to about 5\% of its plain representation, at the price of accessing the graph during expansion (the plain representation expands 23 times faster, even if we use a fast plain graph representation to accompany our encoding). 
\item We introduce \emph{tunnels}: maximal connected subgraphs of $\Hs$ that are regenerated from a stored entry
point by re-expanding the traversal, exploring and discarding the dead alternatives, until reaching another stored entry or a terminal node.
We model the extra work by considering the traversal cost of the graph representaion in use.
\item We bound the extra work per tunnel with an absolute budget $W$ or with a budget
proportional to the output of the tunnel, $c\cdot\mathit{out}+W_0$, which keeps the decoding cost
output-sensitive. We give a bottom-up algorithm to choose what nodes of $\Hs$ to preserve and study how to encode them.
\item We introduce a new representation of paths, orthogonal to tunnelling, for variable-to-variable queries: We cut the expression at a
selective point, following the split strategy of Ring-RPQ~\cite{AGHNR24} or an alternative based
on bridges of the automaton, and represent all solutions as products of leftwards and rightwards traversals, much in the spirit of factorized databases~\cite{OZ15,OS16}. When combining it with tunnels, we also allow tunnelling the generation of the cut points of the graph.
\item We evaluate our techniques on 1,589 real RPQs over a 959M-triple Wikidata
graph~\cite{HRRS19}, and on 1,000 RPQs generated from the expression shapes found in the query
logs~\cite{HM25}, against the explicit $\Hs$, the recomputation from scratch, and, for
variable-to-variable queries, a single PMR for all the sources, on four graph representations.
Tunnels store 14\% of the explicit compressed $\Hs$ of constant-to-variable queries (0.74\% of the plain representation of $\Hs$), while expanding in 0.6--1.3 times the time of
the explicit compressed $\Hs$. For constant-to-constant queries, choosing the endpoint to start from reduces the
traversal 112 times, so that re-running costs about the same as expanding. The factorized representation of variable-to-variable queries takes 8 times less than the PMR of all sources, and tunnelling reduces the space to 0.5\% of the explicit compresed $\Hs$. 
\end{itemize}

\section{Preliminaries}
\label{sec:prelim}

\paragraph{Graphs and RPQs.}
A (labeled) graph is a pair $\G=(V,E)$ with $E\subseteq V\times\Sigma\times V$. For each label $a\in\Sigma$ we also assume the existence of its ``inverse''
$\hat a \in \hat{\Sigma}$, assuming $(v,\hat a,u) \in E$ whenever $(u,a,v)\in E$ (this is done to support query languages, which can express reverse edges). An RPQ is a triple
$(x,r,y)$ where $r$ is a regular expression over $\Sigma\cup\hat\Sigma$, and $x,y$ are variables or
constants in $V$. We call the query \emph{\ctoc} if both $x,y \in V$,
\emph{\ctov} if $x \in V$ and $y$ is a variable, or \emph{\vtov} if both $x$
and $y$ are variables; a \emph{\vtoc} query is handled as a \ctov one by reversing $r$ ($a$ and $\hat{a}$ being mutual reverses).

\paragraph{Automata and product graph.}
We use the Glushkov \cite{Glushkov61} (nondeterministic) automaton $\A=(Q,\Sigma\cup\hat\Sigma,\delta,q_0,F)$ of $r$, which has no
$\varepsilon$-transitions and $|r|+1$ states, where $|r|$ counts only the letters in $r$~\cite{Glushkov61}. The product graph $\G\times\A$ has
nodes $(v,q) \in V \times Q$ and an edge $(v,q)\to(w,q')$ whenever $(v,a,w)\in E$ and $q'\in\delta(q,a)$.
The traversal starts at a \emph{root}: the product node $(s,q_0)$ of the source $s$ in \ctov and
\ctoc queries, or a virtual node whose successors are several such nodes (Section \ref{sec:classes}).
Let $\R$ be the subgraph of the product graph reachable from the root.
The \emph{solution subgraph} $\Hs\subseteq\R$
contains the edges of $\R$ that lie on paths to accepting nodes $(t,f)$, $f\in F$ (with $t$
fixed in \ctoc queries). Every path in $\Hs$ from the root to an accepting node spells a
solution and vice versa. Taking the graph sizes as their number of edges, the cost to compute the RPQ is essentially $O(|\R|)$ and the space to represent its solutions with a PMR~\cite{MNPRVV23} is $O(|\Hs|)$. We say that the nodes in $\R\setminus\Hs$ are \emph{dead}.

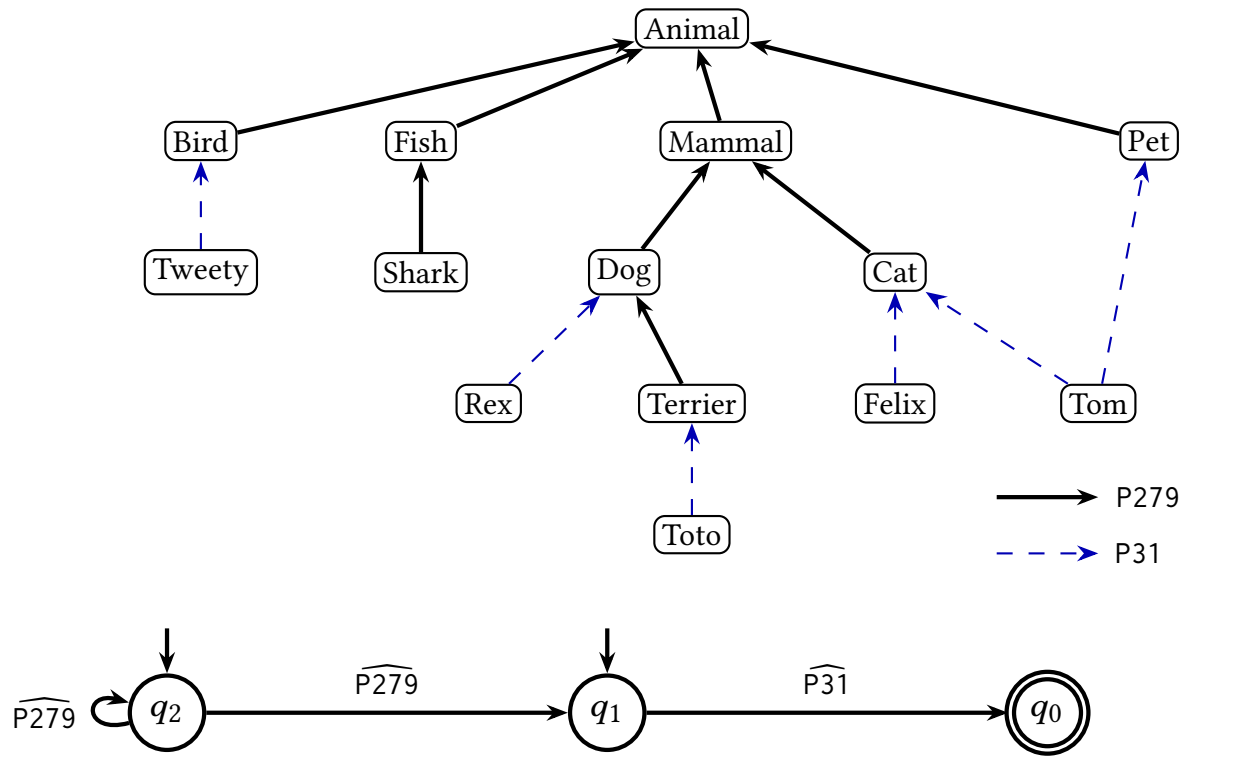
\begin{figure}[t]
\centering
\begin{tikzpicture}[>={Stealth[length=4pt]}, x=1.12cm, y=0.62cm, every node/.style={font=\scriptsize},
  v/.style={rectangle, rounded corners=2pt, draw, inner sep=1.5pt},
  sub/.style={->, thick}, ins/.style={->, dashed, blue!70!black}]
  \node[v] (An) at (3.4,0) {Animal};
  \node[v] (Bi) at (0.5,-1.2) {Bird};   \node[v] (Fi) at (1.8,-1.2) {Fish};
  \node[v] (Ma) at (3.6,-1.2) {Mammal}; \node[v] (Pe) at (6.1,-1.2) {Pet};
  \node[v] (Tw) at (0.5,-2.6) {Tweety}; \node[v] (Sh) at (1.8,-2.6) {Shark};
  \node[v] (Do) at (3.0,-2.6) {Dog};    \node[v] (Ca) at (4.6,-2.6) {Cat};
  \node[v] (Re) at (2.2,-4.0) {Rex};    \node[v] (Te) at (3.4,-4.0) {Terrier};
  \node[v] (Fe) at (4.6,-4.0) {Felix};  \node[v] (To) at (5.8,-4.0) {Tom};
  \node[v] (Tt) at (3.4,-5.4) {Toto};
  \foreach \a/\b in {Bi/An,Fi/An,Ma/An,Pe/An,Sh/Fi,Do/Ma,Ca/Ma,Te/Do} \draw[sub] (\a) -- (\b);
  \foreach \a/\b in {Tw/Bi,Re/Do,Fe/Ca,To/Ca,To/Pe,Tt/Te} \draw[ins] (\a) -- (\b);
  \draw[sub] (5.2,-5.0) -- (5.8,-5.0) node[right] {\texttt{P279}};
  \draw[ins] (5.2,-5.6) -- (5.8,-5.6) node[right, text=black] {\texttt{P31}};
  \begin{scope}[shift={(0.3,-7.3)}, st/.style={circle, draw, thick, minimum size=14pt, inner sep=0pt}]
    \node[st] (q2) at (0,0) {$q_2$};
    \node[st] (q1) at (2.6,0) {$q_1$};
    \node[st, double] (q0) at (5.2,0) {$q_0$};
    \draw[->, thick] (q2) to[loop left, looseness=6] node[left] {$\widehat{\texttt{P279}}$} (q2);
    \draw[->, thick] (q2) -- node[above] {$\widehat{\texttt{P279}}$} (q1);
    \draw[->, thick] (q1) -- node[above] {$\widehat{\texttt{P31}}$} (q0);
    \draw[<-, thick] (q2) -- ++(0,0.9);
    \draw[<-, thick] (q1) -- ++(0,0.9);
  \end{scope}
\end{tikzpicture}
\caption{Running example. Top: a small graph in the style of Wikidata, with edges \texttt{P279}
(subclass of) and \texttt{P31} (instance of). Bottom: the automaton used to evaluate the query
$(x,\texttt{P31/P279*},\textit{Animal})$ backwards from Animal, with initial states $q_1$ and
$q_2$ and final state $q_0$; all the transitions leaving a state have the same label.}
\Description{A class hierarchy rooted at Animal, with subclasses Bird, Fish, Mammal and Pet, then Shark, Dog, Cat and Terrier, and instances Tweety, Rex, Felix, Tom and Toto; Tom is an instance of both Cat and Pet. Below, a three-state automaton.}
\label{fig:runG}
\end{figure}

\paragraph{Running example.}
Figure~\ref{fig:runG} shows a small graph in the style of Wikidata, where \texttt{P279} means
\emph{subclass of} and \texttt{P31} means \emph{instance of}, and the \vtoc query
$(x,\texttt{P31/P279*},\textit{Animal})$, which asks for the instances of Animal or of any of its
(transitive) subclasses. We will evaluate it from Animal, so the traversal follows
the edges backwards, by the inverse labels $\widehat{\texttt{P279}}$ and $\widehat{\texttt{P31}}$, and uses
the Glushkov automaton in reverse: its states $q_1$ and $q_2$ correspond to the letters \texttt{P31}
and \texttt{P279} of the expression, all the transitions leaving each of them carry that letter, and the
initial state $q_0$ of the Glushkov automaton becomes the final one. 

\begin{figure*}[t]
\centering
\begin{tikzpicture}[>={Stealth[length=3.5pt]}, x=3.1cm, y=0.6cm, every node/.style={font=\scriptsize},
  nd/.style={rectangle, rounded corners=2pt, draw, thick, inner sep=1.4pt},
  acc/.style={nd, double},
  dead/.style={rectangle, rounded corners=2pt, draw=black!50, dashed, text=black!65, inner sep=1.4pt},
  e/.style={->, thick}, de/.style={->, draw=black!45, dashed},
  li/.style={pos=0.72, fill=white, inner sep=0.6pt, font=\tiny}]
  \node[nd] (r) at (0,-4.6) {$\rho$};
  \node[nd] (A2) at (1,-4.6) {Animal,$q_2$};
  \node[dead] (A1) at (1,-9.4) {Animal,$q_1$};
  \node[nd] (B1) at (2,0) {Bird,$q_1$};
  \node[dead] (B2) at (2,-1) {Bird,$q_2$};
  \node[dead] (F1) at (2,-2) {Fish,$q_1$};
  \node[dead] (F2) at (2,-3) {Fish,$q_2$};
  \node[dead] (M1) at (2,-4.2) {Mammal,$q_1$};
  \node[nd] (M2) at (2,-5.6) {Mammal,$q_2$};
  \node[dead] (P2) at (2,-7.2) {Pet,$q_2$};
  \node[nd] (P1) at (2,-8.4) {Pet,$q_1$};
  \node[acc] (Tw) at (3,0) {Tweety,$q_0$};
  \node[dead] (S1) at (3,-2.4) {Shark,$q_1$};
  \node[dead] (S2) at (3,-3.3) {Shark,$q_2$};
  \node[nd] (D2) at (3,-4.4) {Dog,$q_2$};
  \node[nd] (D1) at (3,-5.4) {Dog,$q_1$};
  \node[dead] (C2) at (3,-6.4) {Cat,$q_2$};
  \node[nd] (C1) at (3,-7.4) {Cat,$q_1$};
  \node[nd] (T1) at (4,-3.9) {Terrier,$q_1$};
  \node[dead] (T2) at (4,-4.8) {Terrier,$q_2$};
  \node[acc] (Re) at (4,-5.8) {Rex,$q_0$};
  \node[acc] (Fe) at (4,-7.0) {Felix,$q_0$};
  \node[acc] (To) at (4,-8.2) {Tom,$q_0$};
  \node[acc] (Tt) at (5,-3.9) {Toto,$q_0$};
  \draw[de] (r) -- node[li] {0} (A1);   \draw[e] (r) -- node[li] {1} (A2);
  \draw[e] (A2) -- node[li] {0} (B1);   \draw[de] (A2) -- node[li] {1} (B2);
  \draw[de] (A2) -- node[li] {2} (F1);  \draw[de] (A2) -- node[li] {3} (F2);
  \draw[de] (A2) -- node[li] {4} (M1);  \draw[e] (A2) -- node[li] {5} (M2);
  \draw[e] (A2) -- node[li] {6} (P1);   \draw[de] (A2) -- node[li] {7} (P2);
  \draw[e] (B1) -- (Tw);
  \draw[de] (F2) -- (S1); \draw[de] (F2) -- (S2);
  \draw[e] (M2) -- node[li] {0} (C1);   \draw[de] (M2) -- node[li] {1} (C2);
  \draw[e] (M2) -- node[li] {2} (D1);   \draw[e] (M2) -- node[li] {3} (D2);
  \draw[e] (D2) -- node[li] {0} (T1);   \draw[de] (D2) -- node[li] {1} (T2);
  \draw[e] (D1) -- (Re);
  \draw[e] (C1) -- node[li] {0} (Fe);   \draw[e] (C1) -- node[li] {1} (To);
  \draw[e] (P1) -- (To);
  \draw[e] (T1) -- (Tt);
\end{tikzpicture}
\caption{The product graph $\R$ of the running example, explored from the virtual root $\rho$.
Solid nodes and edges form the solution subgraph $\Hs$ (doubled states are accepting nodes), and dashed ones are
dead. Each of the 6 paths from $\rho$ to an accepting node spells a solution; the two paths to Tom
(via Cat and via Pet) share the node (Tom,$q_0$). Small numbers on the edges are the local indices
of the successors, for the nodes with more than one successor. For example, (Animal,$q_2$) has 8
successors: its 4 neighbors by $\widehat{\texttt{P279}}$ (Bird, Fish, Mammal and Pet, in this order) times
the 2 states of $\delta(q_2,\widehat{\texttt{P279}})=\{q_1,q_2\}$.}
\Description{A layered drawing of the 24 nodes of the product graph, from the root on the left to the accepting nodes Tweety, Toto, Rex, Felix and Tom on the right; 14 nodes are solid (the solution subgraph) and 10 are dashed (dead).}
\label{fig:runR}
\end{figure*}
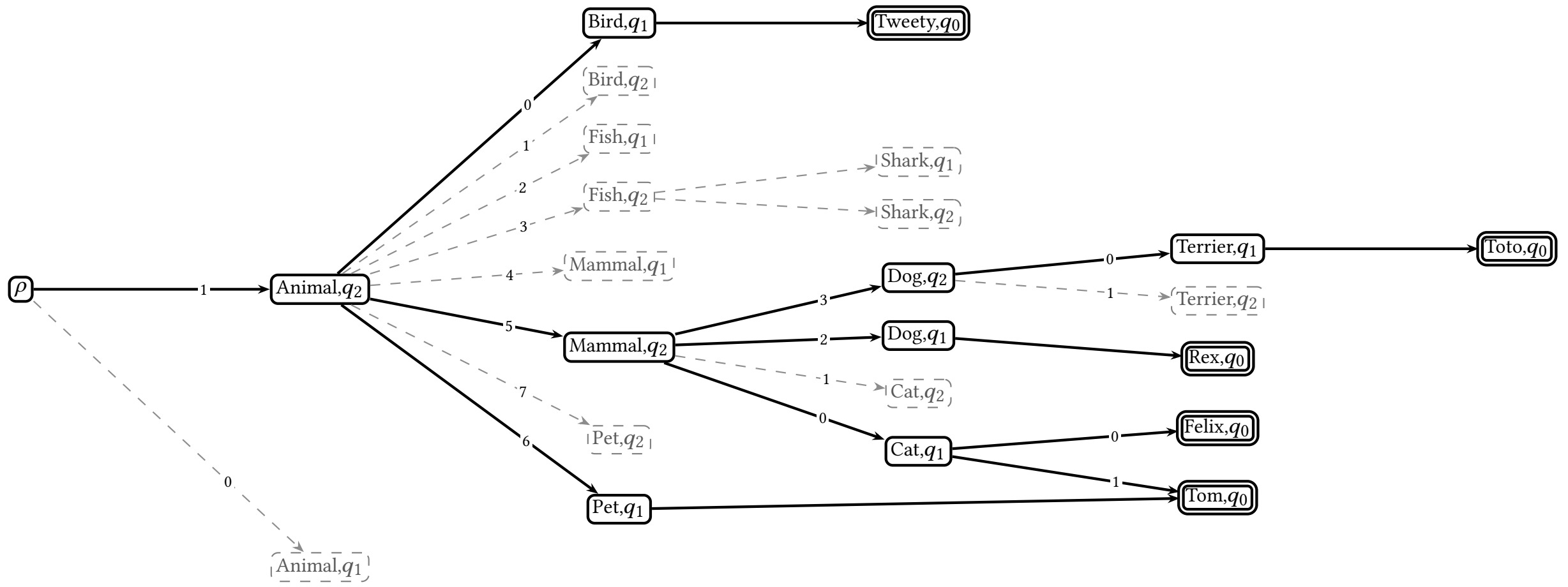

Figure~\ref{fig:runR} shows $\R$. Since a solution can end with
either letter, the root $\rho$ is a virtual node whose successors are (Animal,$q_1$) and
(Animal,$q_2$). The 24 nodes of $\R$ include the 14 of $\Hs$, whose 6
paths from $\rho$ to accepting nodes spell the solutions for the 5 answers (Tom is an instance of Cat and of Pet, so two paths
reach it, sharing the node (Tom,$q_0$)). The other 10 nodes are dead. For example, the traversal
explores (Fish,$q_2$) because Fish is a subclass of Animal, but neither Fish nor its subclass Shark has
instances, and it also explores (Animal,$q_1$), (Bird,$q_2$), and so on, looking for direct instances of Animal
and for subclasses of Bird.

\paragraph{Compact graph representations.}
The Ring~\cite{AHNRRS21} represents the graph as three sequences, containing source nodes, edge labels, and target nodes, in an order that implicitly encodes the edges; each edge is stored in both directions to easily support inverse labels. Its space is close to that of encoding the graph as an array of edges (source, label, target), while it efficiently supports RPQs~\cite{AHNR22,AGHNR24}; we call \emph{Ring-RPQ} the algorithm that evaluates them on the Ring. It can, in $O(\log|V|)$ time, (i) determine if a given node $w$ has a successor by a given label $a$, and (ii) enumerate each such successor.
The representation of Navarro and Robert~\cite{NR24} (which we call Compact) uses less space 
because it encodes the edges in only one direction. It has less functionality than the Ring, but it still supports operations (i) and (ii), and in sublogarithmic time. We complete the set with a plain adjacency
 list representation, which stores both directions with fixed-width identifiers, and is the the largest and fastest. In all the representations, the successors of a node $v$ by the same letter $a$ form a range in the adjacency list of $v$, which we exploit to differentially compress $\Hs$ (see next).

In a Glushkov automaton every transition entering a state has the same label; we use them in reverse so that nodes in the product graph have all their successor nodes by the same letter. This means that the {\em analysis cost} of visiting a node is, independently of the extra complexity factors of each index, one unit of work to determine that it has successors by a symbol, plus one unit to get each of those successors. The native algorithms of Ring and Compact propagate sets of states bit-parallelly, processing several product nodes at once, but we use the graph representations only to support the expansion of
product nodes, independently of how $\Hs$ is computed.

\paragraph{Plain and differential encoding of $\Hs$.} A {\em plain} representation of $\Hs$ numbers its nodes consecutively and stores, for each node, its pair $(v,q)$ (as one integer) and the list of the numbers of its successors $(w,q')$. The label of an edge need not be stored: as we use Glushkov automata in reverse, all the transitions leaving $q$ have the same label. Having the graph $G$ at hand enables an alternative, {\em differential} encoding of $\Hs$: each successor $(w,q')$ by letter $a$ is identified by a \emph{local index}: the rank of $w$ among the neighbors (in $G$) of $v$ by letter $a$ (i.e., an offset in the range), combined with the rank of $q'$ among the states in $\delta(q,a)$. 
We report the size of $\Hs$ in plain and differential form.
Table~\ref{tab:runenc} shows both encodings for the running example. The successors of
(Animal,$q_2$) are its 4 neighbors by $\widehat{\texttt{P279}}$ (Bird, Fish, Mammal and Pet) combined
with the 2 states of $\delta(q_2,\widehat{\texttt{P279}})=\{q_1,q_2\}$, so they have local indices 0
to 7, and the live ones, (Bird,$q_1$), (Mammal,$q_2$) and (Pet,$q_1$), have local indices 0, 5 and 6.
The leaves need not be stored in the differential encoding, since they only appear as targets of the
local indices. The plain encoding spends $\lceil\log_2(|V|\cdot|Q|)\rceil$ bits per key and $\lceil\log_2|\Hs|\rceil$ bits per
successor, whereas the local indices are small numbers that grow within each list, so their gaps are even smaller. In this toy example the difference is small, but on our Wikidata queries the differential encoding is about 19 times smaller than the plain one (Section~\ref{sec:results}).

Decoding a local index $i$ of a node $(v,q)$ with letter $a$ is simple: with $k=|\delta(q,a)|$,
$\lfloor i/k\rfloor$ is the position of $w$ in the range of neighbors of $v$ by $a$, which is accessed
directly in the graph representation, and $i \bmod k$ is the rank of $q'$ in $\delta(q,a)$. For example,
local index 5 of (Animal,$q_2$) gives $\lfloor 5/2\rfloor=2$, the third neighbor (Mammal), and
$5 \bmod 2=1$, the second state ($q_2$), that is, (Mammal,$q_2$).

\begin{table}[t]
\caption{The two encodings of $\Hs$ for the running example (Figure~\ref{fig:runR}), with the
nodes numbered in breadth-first order. The plain encoding stores, for every node, its key $(v,q)$ and
the numbers of its successors. The differential encoding stores only the keys of the internal nodes,
each with the local indices of its live successors, gap-encoded.}
\label{tab:runenc}
\centering\footnotesize
\begin{tabular}{@{}rlll@{}}
\toprule
 & & \multicolumn{1}{c}{plain} & \multicolumn{1}{c}{differential} \\
\# & node $(v,q)$ & successors & local indices \\
\midrule
0 & $\rho$ & 1 & 1 \\
1 & (Animal,$q_2$) & 2, 3, 4 & 0, 5, 6 \\
2 & (Bird,$q_1$) & 5 & 0 \\
3 & (Mammal,$q_2$) & 6, 7, 8 & 0, 2, 3 \\
4 & (Pet,$q_1$) & 9 & 0 \\
5 & (Tweety,$q_0$) & -- & -- \\
6 & (Cat,$q_1$) & 10, 9 & 0, 1 \\
7 & (Dog,$q_1$) & 11 & 0 \\
8 & (Dog,$q_2$) & 12 & 0 \\
9 & (Tom,$q_0$) & -- & -- \\
10 & (Felix,$q_0$) & -- & -- \\
11 & (Rex,$q_0$) & -- & -- \\
12 & (Terrier,$q_1$) & 13 & 0 \\
13 & (Toto,$q_0$) & -- & -- \\
\bottomrule
\end{tabular}
\end{table}

\section{Tunnels}
\label{sec:tunnels}

\paragraph{Model.}
We assume that the decoder has access to the graph (represented as a Ring or in another form), to the RPQ, and to a compact structure $\Sset$ storing a {\em sample} of the nodes in $\Hs$. Its task is to {\em simulate} access to $\Hs$ (for example, to traverse its paths):
If a node $x \in \Hs$ is not represented in $S$, then traversing it requires the decoder to recompute its successors in $\Hs$ transitively until reaching the next nodes described in $S$ or terminal nodes (with no further successors). In addition, it must report any node $(t,f) \in \Hs$ with $f \in F$ it hits. Along this expansion,
a dead successor $y$ cannot be told apart without exploring its whole forward closure
$\mathit{Cl}(y)\subseteq\R\setminus\Hs$. Hence the {\em extra work} to re-expand $x$ is
\[ \cost(x) \;=\; \sum_{y\ \text{dead successor of}\ x}\ \ \sum_{z=(v,q)\in\mathit{Cl}(y)} \big(1+\an(z)\big), \]
where $1+\an(z)$ is the analysis cost of Section~\ref{sec:prelim}:
the 1 accounts for producing $v$ from the range of its parent (operation (ii)), and $\an(z)$ is 1 if $q$ has outgoing transitions, for checking
whether $v$ has successors by their label (operation (i), only 1 label as we use Glushkov in reverse direction), and 0 otherwise. A dead leaf $z=(v,q)$ (with no successors)
thus usually costs 2: it is produced, and then found to have no successor by the label of the edges leaving $q$. Nodes of $\Hs$ with no dead
successor ($\cost(x)=0$) are free to regenerate.
In the running example (Figure~\ref{fig:runR}), $\cost(\rho)=2$, for the dead leaf (Animal,$q_1$);
$\cost(\text{Mammal},q_2)=2$, for (Cat,$q_2$); $\cost(\text{Dog},q_2)=2$, for (Terrier,$q_2$); and
$\cost(\text{Animal},q_2)=14$: its dead successors are four leaves of cost 2 and (Fish,$q_2$), whose closure
also contains (Shark,$q_1$) and (Shark,$q_2$), so it costs 6. All the other nodes of $\Hs$ cost 0.

\paragraph{Representation.}
We store a set $\Sset\subseteq\Hs$ of nodes, each with the list of its live successors. Alternatively,
since all successors can be enumerated, a stored node may keep the list of its \emph{dead}
successors, which are skipped; we keep the shorter of the two. All other nodes of $\Hs$ are
re-expanded when needed. In the running example, (Mammal,$q_2$) has 3 live successors and 1 dead one, so it would store its dead list, $\langle 1\rangle$. The decoder recognizes stored nodes through a dictionary on product nodes.
Neither labels nor automaton states are stored: a successor is identified by its position among the
successors that the expansion of its parent produces, from which the decoder recovers the
neighbor, the label and the state. Leaves of
$\Hs$ cost nothing in an explicit representation of $\Hs$, as they only appear as edge targets, and
they are never stored in $\Sset$ either, so we do not count them in either of the two representations.

We also use a differential encoding for $S$: The product nodes of $\Sset$ are kept as a sorted list of {\em global} identifiers of $\Hs$, compacted with 
Elias--Fano~\cite{Eli74,Fan71,OS07}, which answers membership and rank. Each stored node of $\Sset$ then keeps its list of (live or dead) {\em local} indices, in increasing order and gap-encoded with $\delta$-codes. 

\begin{figure*}[t]
\claudecolor
\centering
\newcommand{\drawRunH}{%
  \node[nd,Sr] (r) at (3.4,0) {$\rho$};             \node[cst, right=1pt of r] {2};
  \node[nd,SA2] (A2) at (3.4,-1) {Animal,$q_2$};    \node[cst, right=1pt of A2] {14};
  \node[nd,SB1] (B1) at (0.5,-2) {Bird,$q_1$};
  \node[nd,SM2] (M2) at (3.4,-2) {Mammal,$q_2$};    \node[cst, right=1pt of M2] {2};
  \node[nd,SP1] (P1) at (6.5,-2) {Pet,$q_1$};
  \node[acc,SB1] (Tw) at (0.5,-3) {Tweety,$q_0$};
  \node[nd,SD2] (D2) at (2.0,-3) {Dog,$q_2$};       \node[cst, above left=-1pt and 0pt of D2] {2};
  \node[nd,SC1] (D1) at (3.4,-3) {Dog,$q_1$};
  \node[nd,SC1] (C1) at (4.8,-3) {Cat,$q_1$};
  \node[nd,ST1] (T1n) at (2.0,-4) {Terrier,$q_1$};
  \node[acc,SC1] (Re) at (3.4,-4) {Rex,$q_0$};
  \node[acc,SC1] (Fe) at (4.8,-4) {Felix,$q_0$};
  \node[acc,STo] (To) at (6.3,-4) {Tom,$q_0$};
  \node[acc,ST1] (Tt) at (2.0,-5) {Toto,$q_0$};
  \draw[e] (r) -- (A2); \draw[e] (A2) -- (B1); \draw[e] (A2) -- (M2); \draw[e] (A2) -- (P1);
  \draw[e] (B1) -- (Tw); \draw[e] (M2) -- (D2); \draw[e] (M2) -- (D1); \draw[e] (M2) -- (C1);
  \draw[e] (P1) -- (To); \draw[e] (C1) -- (To); \draw[e] (C1) -- (Fe); \draw[e] (D1) -- (Re);
  \draw[e] (D2) -- (T1n); \draw[e] (T1n) -- (Tt);}
\begin{tikzpicture}[>={Stealth[length=3.5pt]}, x=1cm, y=0.82cm, every node/.style={font=\scriptsize},
  nd/.style={rectangle, rounded corners=2pt, draw, inner sep=1.4pt},
  acc/.style={nd, double},
  st/.style={rounded corners=0pt, ultra thick},
  T0/.style={fill=blue!20}, T1/.style={fill=orange!35}, T2/.style={fill=green!35}, T3/.style={fill=cyan!30},
  cst/.style={font=\scriptsize\itshape, text=red!70!black},
  e/.style={->, thick},
  ttl/.style={font=\small}]
  \begin{scope}[Sr/.style={T0}, SA2/.style={st,T1}, SB1/.style={T1}, SM2/.style={T1}, SP1/.style={T1},
                SD2/.style={T1}, SC1/.style={T1}, ST1/.style={T1}, STo/.style={T1}]
    \drawRunH
    \node[ttl] at (0.9,0) {$W=4$};
  \end{scope}
  \begin{scope}[xshift=9.2cm, Sr/.style={st}, SA2/.style={st,T1}, SB1/.style={T1}, SM2/.style={st,T2},
                SP1/.style={T1}, SD2/.style={st,T3}, SC1/.style={T2}, ST1/.style={T3},
                STo/.style={path picture={\fill[orange!35] (path picture bounding box.south west) rectangle (path picture bounding box.north);
                  \fill[green!35] (path picture bounding box.north) rectangle (path picture bounding box.south east);}}]
    \drawRunH
    \node[ttl] at (0.9,0) {$W=0$};
  \end{scope}
\end{tikzpicture}
\caption{Tunnels on the running example: two copies of the solution subgraph $\Hs$ of Figure~\ref{fig:runR},
with $\cost(x)$ in red next to the nodes with dead successors. Stored nodes have a thick square border,
and each tunnel is shaded in a different color, which is also used for its entry. Left: with $W=4$ we store only (Animal,$q_2$),
with its live list $\langle 0,5,6\rangle$, which leaves two tunnels: the one entered at the root (blue), with extra work
$\cost(\rho)=2$, and the one entered at (Animal,$q_2$) (orange), with extra work $2+2=4$.
Right: with $W=0$ we store the 4 nodes with dead successors: $\rho$ with $\langle 1\rangle$, (Animal,$q_2$)
with $\langle 0,5,6\rangle$, (Mammal,$q_2$) with its dead list $\langle 1\rangle$, and (Dog,$q_2$) with $\langle 0\rangle$.
The tunnel of the root is empty, the other three (orange, green and cyan) have no extra work, and (Tom,$q_0$) belongs to two of them,
those entered at (Animal,$q_2$) and (Mammal,$q_2$).
With $W\ge 20$, or with the relative budget $4\cdot\mathit{out}+10$, nothing is stored and the decoder re-runs the query.}
\Description{Two copies of the 14 nodes of the solution subgraph drawn top-down from the root. On the left, for W=4, the root is shaded blue, all the other nodes orange, and only the node (Animal, q2) is stored. On the right, for W=0, the root, (Animal, q2), (Mammal, q2) and (Dog, q2) are stored and start orange, green and cyan tunnels; the node Tom is shaded half orange and half green.}
\label{fig:example}
\end{figure*}
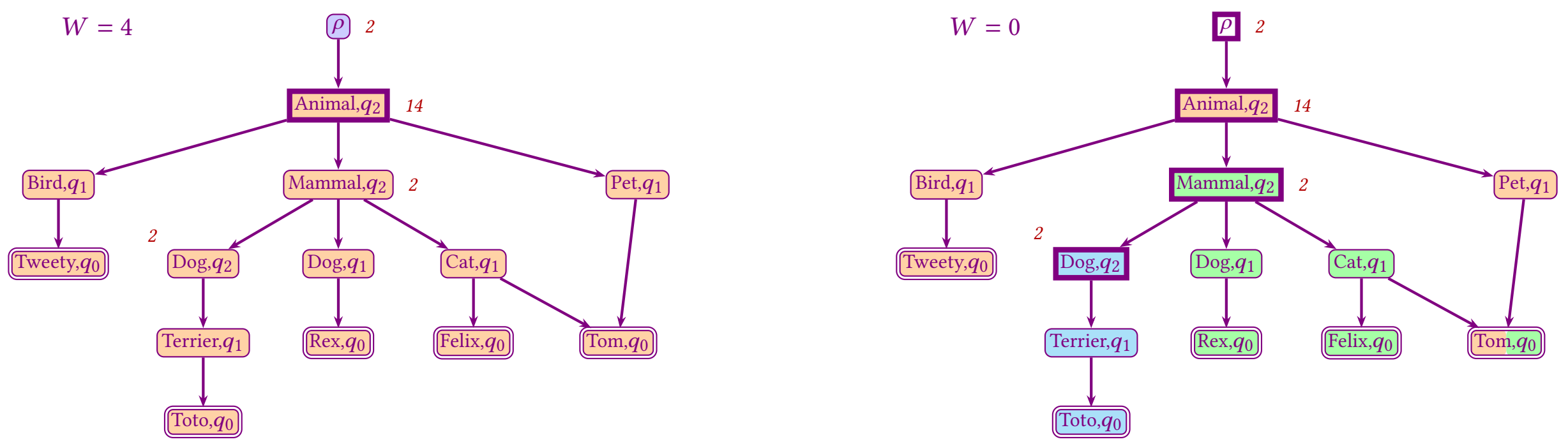

\paragraph{Tunnels and budgets.}
A \emph{tunnel} is the set of non-stored nodes (i.e., not in $S$) reachable from the root or from a
stored node (its \emph{entry}) without crossing other stored nodes. Each tunnel has a single
entry, but tunnels may overlap: a non-stored node reachable from several entries belongs to the
tunnel of each (Figure~\ref{fig:example}). Its {\em extra work} is the sum of $\cost(x)$ over its
nodes $x$. We bound the allowed extra work of our tunnels with one of two budgets:
\begin{itemize}
\item \emph{absolute:} the extra work of every tunnel is at most $W$;
\item \emph{relative:} the extra work of every tunnel is at most $c\cdot\mathit{out}+W_0$, where
$\mathit{out}$ is the number of edges of $\Hs$ the tunnel contains.
\end{itemize}
Note that $W=0$ stores exactly the nodes with dead successors, so it saves space essentially for free in terms of expansion time. In the other extreme, $W=\infty$ stores nothing, which amounts to re-running the whole query.
Figure~\ref{fig:example} shows the tunnels of the running example for some budgets.

The absolute budget bounds the work to regenerate any single tunnel, independently of what it
produces, which matters for local access: entering a tunnel costs $O(W)$ operations on top of
the edges it outputs. When paths are enumerated, this bounds the extra delay between consecutive
edges output (e.g., by $O(W\log|V|)$ on the Ring), as dead branches are explored only between them. With the relative
budget, instead, enumerating the paths of a tunnel costs $O((1+c)\cdot\mathit{out}+W_0)$ operations,
so the representation is output-sensitive (i.e., it pays an extra cost per delivered edge), but a tunnel with large output may contain
long dead explorations, so the delay is not bounded. Experimentally the relative budget compresses
much more for a similar total expansion time (Section~\ref{sec:results}). 

\paragraph{Choosing what to store.}
On trees with node weights and an absolute bound, partitioning the tree into connected parts of
weight at most $W$ with the fewest cuts is solved optimally by the greedy algorithm of Kundu and
Misra~\cite{KM77}: process nodes bottom-up and, when a subtree exceeds the bound, cut its
heaviest child subtrees. $\Hs$ is a general graph, so we apply this greedy algorithm as a heuristic
on the DAG of its strongly connected components, sinks first. For a component $C$ we compute the
work $w(C)$ of the tunnel it would start: the costs of its non-stored nodes plus the work of the
child components that are not cut. With the absolute budget, the \emph{excess} of a child
component $D$ is $w(D)$; with the relative one it is $w(D)-c\cdot\mathit{out}(D)$, where
$\mathit{out}(D)$ is the output of the tunnel started by $D$, so a child that produces much output
relative to its work does not count against its parent. If $C$ alone exceeds the budget, we store its
nodes of largest cost until it fits. Then we cut its child components, largest excess first, by
storing their entry nodes, until $C$ plus its remaining children fits within the budget.

A component reachable from several parents belongs to the tunnel of each of them, and its work is
counted in each of them, so that none of them exceeds the budget. Its output, instead, is credited
to a single parent, the one through which it was first discovered in the breadth-first traversal of
the construction (any spanning forest would do); otherwise the same edges would pay for the work
of several tunnels. This makes the work an upper bound and the output a lower bound, so the
budgets hold. Within the same parent, however, summing the work of its children counts a shared
descendant once per child, although the decoder expands it only once, since it keeps the visited
nodes. This overestimate can force unnecessary cuts in DAGs where many children share a costly
descendant. We therefore compare the result with two other stored sets that also respect the
budget, and keep the one with the smallest encoding: the set for $W=0$ (every node with dead
successors), and the result of the greedy algorithm when the costly shared nodes (with dead
successors and several parents in $\Hs$) are stored beforehand.

Finally, computing $\cost(x)$ requires exploring dead closures at construction time. We cap the
exploration of a single dead branch (10,000 units in our experiments) to bound the construction
time; a node with a branch reaching the cap is always stored.

\paragraph{Decoding.}
We consider three operations on the representation.
(1) \emph{Rebuilding} $\Hs$, Algorithm~\ref{alg:decode}: a breadth-first traversal from the root
keeps the visited product nodes. At a stored node it reads its list: for a live list it accesses
the listed neighbors directly; for a dead list it enumerates all successors and skips the listed
ones. At a non-stored node it expands all the transitions. Finally, a backward pass from the
accepting nodes discards the dead nodes explored.
(2) \emph{Expanding a single tunnel} from its entry, Algorithm~\ref{alg:decode-tunnel}, as needed for
local access: the traversal stops at stored nodes, which belong to $\Hs$ by definition, so the
backward pass starts from them as well as from the accepting nodes; its cost is the output of the
tunnel plus its extra work, which the budget bounds. When rebuilding all of $\Hs$, we retain the
expanded nodes, so a node shared by several tunnels is
expanded only once.
(3) \emph{Traversing} $\Hs$ without rebuilding it, for example to enumerate its paths with the usual
depth-first traversal of a PMR~\cite{MNPRVV23}: the traversal can expand each tunnel as it enters
it, with operation (2), and then continue on the expanded tunnel until reaching a stored node.
Integrating the traversal with the expansion, as a full traversal guided by hints, would avoid
materializing the tunnel, but the traversal could not tell dead branches apart without exploring
them, and it would explore them again every time it enters the tunnel.
When $\Hs$ has cycles,
enumerating paths also needs a path semantics (e.g., simple paths or trails) to terminate, which is
orthogonal to tunnelling.

\begin{algorithm}[t]
\caption{Decoding $\Hs$ from $\Sset$. Nodes are pairs $(w,q) \in V \times Q$. $P$ is a queue of pending nodes. $\mathsf{Succ}(x,i)$ returns the $i$th successor of node $x$ from the graph $G$.}
\label{alg:decode}
\small
\DontPrintSemicolon
\SetKwFunction{Succ}{Succ}\SetKwProg{Fn}{Function}{:}{}
\KwIn{the root $\rho$, the stored set $\Sset$ with the list $L(x)$ of each $x\in\Sset$, and the graph}
\KwOut{the nodes and edges of $\Hs$}
$V' \gets \{\rho\}$; $E' \gets \emptyset$; $P \gets [\rho]$ \tcp*{forward}
\While{$P$ is not empty}{
  $x \gets$ dequeue$(P)$; $n \gets$ number of successors of $x$\;
  \lIf{$x\in\Sset$ with a live list}{$I \gets L(x)$ \tcp*[f]{direct}}
  \lElseIf{$x\in\Sset$ with a dead list}{$I \gets [1,n]\setminus L(x)$}
  \lElse{$I \gets [1,n]$ \tcp*[f]{re-expand}}
  \ForEach{$i\in I$}{
    $y\gets$ \Succ{$x$, $i$}; $E' \gets E'\cup\{(x,y)\}$\;
    \lIf{$y\notin V'$}{$V'\gets V'\cup\{y\}$; enqueue$(P,y)$}
  }
}
$V'' \gets \{(t,f)\in V' : f\in F\}$; $E''\gets\emptyset$; $P \gets V''$ \tcp*{backward}
\While{$P$ is not empty}{
  $y \gets$ dequeue$(P)$\;
  \ForEach{$(x,y)\in E'$}{$E'' \gets E'' \cup\{(x,y)\}$\;
    \lIf{$x\notin V''$}{$V''\gets V''\cup\{x\}$; enqueue$(P,x)$}}
}
\Return $(V'',E'')$\;
\end{algorithm}

\begin{algorithm}[t]
\caption{Decoding the tunnel of entry point $\rho$. Notation is as in Algorithm~\ref{alg:decode}.}
\label{alg:decode-tunnel}
\small
\DontPrintSemicolon
\SetKwFunction{Succ}{Succ}\SetKwProg{Fn}{Function}{:}{}
\KwIn{the entry point $\rho$, its number $n$ of successors and its list $L(\rho)$, the stored set $\Sset$, and the graph}
\KwOut{the nodes and edges of the tunnel of $\Hs$ with entry point $\rho$}
$V' \gets \{\rho\}$; $E' \gets \emptyset$; $P \gets [\,]$\;
$I \gets L(\rho)$ if it is a live list, else $[1,n]\setminus L(\rho)$\;
\ForEach{$i\in I$}{
  $y \gets$ \Succ{$\rho$, $i$}; $E' \gets E'\cup\{(\rho,y)\}$\;
  \lIf{$y\notin V'$}{$V'\gets V'\cup\{y\}$; \textbf{if} $y\notin\Sset$ \textbf{then} enqueue$(P,y)$}
}
\While{$P$ is not empty}{
  $x \gets$ dequeue$(P)$\;
  \ForEach{$y\in \Succ{$x$}$}{
    $E' \gets E'\cup\{(x,y)\}$\;
    \If{$y\notin V'$}{$V'\gets V'\cup\{y\}$; \textbf{if} $y\notin\Sset$ \textbf{then} enqueue$(P,y)$}
  }
}
$V'' \gets \{(t,f)\in V' : f\in F\} \cup (V'\cap\Sset)$; $E''\gets\emptyset$; $P \gets V''$\;
\While{$P$ is not empty}{
  $y \gets$ dequeue$(P)$\;
  \ForEach{$(x,y)\in E'$}{$E'' \gets E'' \cup\{(x,y)\}$\;
    \lIf{$x\notin V''$}{$V''\gets V''\cup\{x\}$; enqueue$(P,x)$}}
}
\Return $(V'',E'')$\;
\end{algorithm}

\section{The Three Classes of Queries}
\label{sec:classes}

\paragraph{Constant-to-variable.}
In \ctov and \vtoc queries with constant $c$, the only root is $(c,q_0)$ (or, with the reversed automaton, a virtual node whose successors are the nodes $(c,q)$ of its initial states $q$, as in the running example), and the traversal goes forward for \ctov queries
or backward (i.e., forwards with the inverted expression) for \vtoc queries.
As in Ring-RPQ, we actually build the Glushkov automaton of the reversed expression and run it backwards.
The natural case is thus \vtoc, which traverses the edges backwards from the constant, and a \ctov query
is solved as the \vtoc query of its reversed expression.

\paragraph{Constant-to-constant.}
Both endpoints are fixed in \ctoc queries, so $\Hs$ only keeps the nodes leading to the target $t$, and $\R\setminus\Hs$
can be much larger than $\Hs$. Since either endpoint can be the root, the representation can be
built from the side with less dead work.

\paragraph{Variable-to-variable.}
Without constants, a PMR can be built for all the sources at once \cite{MNPRVV23}: the product graph explored
from all of them, with the live sources marked. This shares the structure that solving each source
separately would repeat (in the query \texttt{?x P31/P279* ?y}, a million instances climb the same
class hierarchy). We instead cut the automaton at a point with few \emph{seeds} (occurrences in $G$), evaluate the two
parts leftwards and rightwards from each of the seeds as the source, and represent the solutions as (non-expanded) Cartesian products:
\[ \biguplus_{\text{seed } m} \textit{leftward-paths}(m)\times\textit{rightward-paths}(m). \]
While this way of evaluating \vtov RPQs was already shown to be effective~\cite{AGHNR24}, using it to {\em represent the result} is new, and akin to factorized databases~\cite{OZ15,OS16}.
We study two ways of cutting.
\begin{description}
\item[\emph{Ring-RPQ split}~\cite{AGHNR24}:] The ``mandatory'' symbols (of transitions used by every accepting path of the automaton), are identified. If $\A$ has a pair $a,b$ of consecutive mandatory symbols, then the automaton can be cut at the connecting state and the seeds are the nodes of $G$ that are targets of edges by $a$ and sources of edges by $b$. The best seeds are chosen with a selectivity estimate. 
This split needs \emph{RingAB}, an extended Ring index with structures to estimate selectivities
and intersect ranges of neighbors~\cite{AGHNR24}.
\item[\emph{Bridge split}:] The same mandatory labels are bridges of $\A$.
A simpler strategy is thus to cut at the source or the target of one of them, choosing the cut with the fewest
distinct seed nodes, which the Ring counts with one range query. This split runs on the
basic Ring index, and on any representation that can list the sources and targets of a label.
\end{description}

The automaton $\A$ is cut into a left part, $\A_L$, with reversed edges, and a right part, $\A_R$. Each part is evaluated from a virtual root whose successors are the product nodes of the seeds, forwards or backwards.
As in Ring-RPQ, the first
side is evaluated from all the candidate seeds and the second only from those that have solutions
in the first; for the Ring-RPQ split we use its choice of first side, and for the bridge split we
start with the side that does not begin with the cut label. Both resulting subgraphs, restricted
to the seeds alive on both sides, are compressed with tunnels. The list of live seeds is also
regenerable: the candidate seeds are given by the index, and the decoder evaluates the sides in
the same order. A dead seed that is left for the decoder costs its dead closure in the first side
if it dies there, or else the work it adds in the first side outside $\Hs$ plus its dead closure in
the second side. Within the budget, the cheapest dead seeds are explored and the rest are stored as
an exclusion list, or the live seeds are listed, whichever is shorter.

Both the factorized representation and the PMR of all sources contain essentially the same
product nodes, but they differ in which of them are leaves. The PMR of all sources is traversed
from the sources, which become internal nodes, each with its list of successors. The factorized
representation is traversed from the seeds, backwards on the left part and forwards on the right
one, so that both the sources and the targets are leaves, which cost nothing in a differential
encoding of $\Hs$, and only the seeds, usually far fewer, are roots. The factorized representation is also faster to build and to
re-run, since it explores the graph only from the seeds. The PMR of all sources, on the other hand,
needs no cut, and tunnels can avoid storing its sources.

\section{Experimental setup}
\label{sec:setup}

\paragraph{Data.}
We use the Wikidata subgraph of Arroyuelo et al.~\cite{AHNR22} (the Wikidata Core graph of Hogan et al.~\cite{HRRS19}: the truthy dump of November 2018 without its
multilingual labels and descriptions), featuring 958,844,164 edges 
with 5,419 distinct labels. The logged queries use 143 of those labels.

\paragraph{Queries.}
We use the 1,589 RPQs of Arroyuelo et al.~\cite{AHNR22}, extracted from the Wikidata query
logs~\cite{BMT19,MalyshevEtAl18}. Of them, 1,495 have one constant (\ctov or \vtoc), 6 have two,
and 88 are \vtov. Inverse labels are written in SPARQL syntax (\texttt{\^{}a}, \texttt{\^{}(\dots)}); we
rewrite them into the syntax of Ring-RPQ, which inverts only labels, by pushing the inversion
down ($\hat{\ }(r_1/r_2) = \hat{\ }r_2/\hat{\ }r_1$).
We generate more \ctoc queries from each query with one constant by taking 5 of its answers at random as
targets, which yields 7,253 \ctoc queries. We also drop the constants of queries to obtain more of type \vtov. Many \ctov queries differ only in their
constant, however, so dropping the constants and adding the \vtov queries of the log yields only 299
distinct \vtov queries; 171 of them have a mandatory label to cut at. To obtain a more varied workload, we also generate 1,000 queries from the
expression shapes that Hammerer and Martens found in the Wikidata logs~\cite{HM25}: we choose a
shape with probability proportional to the number of distinct logged RPQs that have it, and bind
its letters to distinct labels with a random walk on the graph that follows a random word of the
expression; each letter takes a label that the current node has, chosen with the frequency of the
label in the query log. The walk guarantees that the query has answers; half of the queries are
\ctov and half \vtoc. The 1,000 queries use 54 shapes, led by \texttt{a/b*} (364), \texttt{a*} (181) and
\texttt{a/b} (133), and dropping their constants yields 440 distinct \vtov queries, 255 of them with a cut.
Totals include all the queries; medians are reported over queries whose $\Hs$ has at least 1,000
edges, below which the space saved is irrelevant in absolute terms.

From the generated queries we derive \ctoc and \vtov queries as from the log: 5 random answers
of each query give 3,696 \ctoc queries, and dropping the constants gives the 440 \vtov queries.

\paragraph{Measurements.}
Every expansion is repeated 3 times (once for \vtov) and the minimum time is reported.
We ran the experiments on a machine with two Intel Xeon Silver 4110 processors at 2.1\,GHz (32
cores in total) and 735\,GB of RAM, running Linux 6.12. The code is in C++, compiled with g++ 14.2
and \texttt{-O3}. All the implementations, ours and the baselines, are sequential: each run uses a
single thread, and several independent runs shared the machine.

\paragraph{Baselines and comparisons.}
For each query, we build $\Hs$ and $\Sset$, with various parameter choices, and compare (i)~traversing the explicit $\Hs$, with plain or differential encoding; (ii)~traversing $\Hs$ having only $\Sset$, via tunnelling; (iii)~fully re-running the query on the graph ($W=\infty$). For \vtov, we also consider the PMR of all sources, whose root keeps the list of live
(or dead) sources, combined with (i)--(iii). 
For each variant we report the time to build $\Sset$ (traversing $\R$, computing $\Hs$ and the costs, and choosing and encoding $\Sset$), its space, and the
time to expand it (i.e., rebuild $\Hs$ with all its edges, Algorithm~\ref{alg:decode}). $\Hs$ and $\Sset$, and thus their
space, are the same on every graph representation, since our cost model is; the times to build and to expand them depend on the representation.

\paragraph{Implementation.}
We implement $\Hs$ and $\Sset$ independently of the graph representation, and evaluate tunnelling on four of them
(Table~\ref{tab:indexes}): the basic \emph{Ring} index of Ring-RPQ~\cite{AGHNR24}; \emph{RingAB}, which
adds the structures of the Ring-RPQ split, the only one where that split runs; \emph{Compact}~\cite{NR24}, which stores each edge once and navigates it in
both directions; and \emph{Plain}, adjacency lists in both directions with 32-bit node and label
identifiers. All four produce the same product graph and the same $\Hs$, and differ in the cost of
expanding a node.

\begin{table}[t]
\caption{The graph representations: space in bytes per edge, construction time, and peak memory
during construction.}
\label{tab:indexes}
\small
\begin{tabular}{lrrr}
\toprule
representation & bytes/edge & construction & memory \\
\midrule
Ring~\cite{AGHNR24} & 16.3 & 1.2 h & 70 GB \\
RingAB~\cite{AGHNR24} & 27.8 & 2.4 h & 103 GB \\
Compact~\cite{NR24} & 7.0 & 0.6 h & 45 GB \\
Plain & 20.9 & 0.1 h & 34 GB \\
\bottomrule
\end{tabular}
\end{table}


\section{Results}
\label{sec:results}

\subsection{Construction}

Table~\ref{tab:hsizes} summarizes the sizes of the solution subgraphs of each class, and the
space of the representations of $\Hs$ that we compare. 
It can be seen that the differential compression reduces $\Hs$ to just 5.2\% of its plain representation on \ctov queries, and to 2.7\% on \vtov, because most nodes have few children, and leaves need no entry of their own. The choice of tunnelling we show further reduces the space of the compressed $\Hs$ to 14\% in \ctov queries and 13\% in \vtov: nearly half of the large queries store at most 1\%
of $\Hs$. On \ctoc, tunnelling reduces the size of plain $\Hs$ to 2\%. Finally, although the all-sources graph of \vtov queries is similar to that of our factorized \vtov representation, it must represent many more edges because sources and targets are leaves in our representation. Interestingly, although the size of the differential representation of $\Hs$ with factorization is 12\% of that of all-sources, tunnelling compresses the latter much better, making it 3 times smaller than that of our factorization.

\begin{table}[t]
\caption{Average size of $\Hs$ per query of each class, in thousands of nodes and edges, and space in KB of
the plain $\Hs$, the differentially encoded $\Hs$, and of $\Sset$
with the relative budget $4\cdot\mathit{out}+10$ on the Ring (RingAB for \vtov).}
\label{tab:hsizes}
\small
\setlength{\tabcolsep}{4pt}
\begin{tabular}{lrrrrr}
\toprule
class & nodes & edges & plain $\Hs$ & diff $\Hs$ & $\Sset$ \\
\midrule
{\ctov} (1,495) & {79.2} & {91.0} & {702.9} & {36.8} & {5.2} \\
{\ctoc} (7,253) & {0.9} & {4.2} & {14.6} & {5.4} & {0.3} \\
{\vtov} (171) & {7,764} & {8,365} & {80,271} & {2,166} & {282} \\
{\vtov, all sources} & {7,698} & {19,242} & {115,862} & {17,790} & {93} \\
\bottomrule
\end{tabular}
\end{table}

Table~\ref{tab:build} shows construction times for $\Hs$ and $\Sset$. Most of the time is spent in the traversal of $\R$, though the time to find the tunnelled representation is not negligible (on Ring and Compact, 20\%--28\% on \ctov and 19\%--43\% in most cases of \vtov). Building the factorized \vtov representation turns out to cost an order of magnitude more on Compact, because it is forced to traverse the graph in both directions, and it is much slower in one direction than in the other (in other cases, it can choose the direction by reversing the expression).

\begin{table}[t]
\caption{Average construction time per query: traversing $\R$, computing $\Hs$ (the backward
pass), and computing $\Sset$ with $4\cdot\mathit{out}+10$,
in milliseconds (seconds for \vtov).}
\label{tab:build}
\small
\setlength{\tabcolsep}{4pt}
\begin{tabular}{llrrr}
\toprule
class & representation & $\R$ & $\Hs$ & $\Sset$ \\
\midrule
\ctov & Ring & 425 & 7.5 & 85 \\
      & Compact & 308 & 8.3 & 87 \\
      & Plain & 51 & 8.1 & 84 \\
\midrule
\ctoc & Ring & 438 & 5.7 & 11 \\
      & Compact & 318 & 6.1 & 12 \\
      & Plain & 52 & 6.2 & 12 \\
\midrule
\vtov & RingAB, Ring-RPQ split & 19.8 & 0.56 & 6.9 \\
      & Ring, bridge split & 35.5 & 0.60 & 6.6 \\
      & Compact, bridge split & 210.7 & 0.65 & 7.3 \\
      & Plain, bridge split & 17.6 & 0.84 & 7.5 \\
      & Ring, all sources & 29.0 & 0.82 & 13.9 \\
\bottomrule
\end{tabular}
\end{table}

\subsection{Constant-to-variable queries}
\label{sec:ctov}

\begin{figure}[t]
\includegraphics[width=\linewidth]{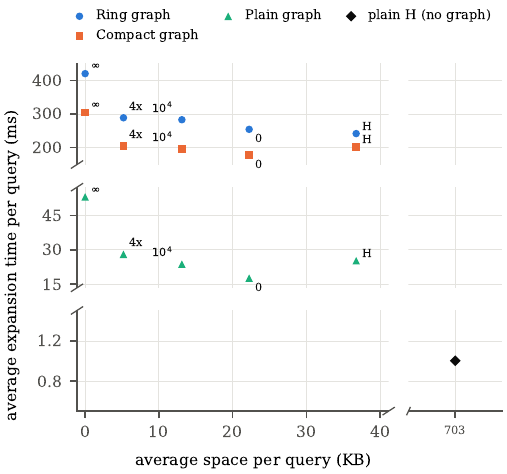}
\caption{Constant-to-variable queries: average space of $\Sset$ per query against average time to expand it on
three representations. Each point is labeled with its budget: $W=0$, $W=10^4$, the relative budget
$4\cdot\mathit{out}+10$ (``4x''), and $W=\infty$, which stores nothing; ``H'' is the compressed $\Hs$, and the diamond
is the plain $\Hs$, which is independent on the graph. Both axes are linear and broken at some points. }
\Description{Scatter plot of average space per query against average expansion time per query for constant-to-variable queries, on the Ring, Compact and Plain graph representations, with one point per budget, the compressed solution subgraph, and the plain solution subgraph.}
\label{fig:ctov}
\end{figure}

\begin{table}[t]
\caption{Constant-to-variable queries on the Ring: space of $\Sset$ relative to the explicit $\Hs$,
median over the {890} queries with $|\Hs|\ge 1000$ edges, fraction of those
queries storing at most 1\%, and expansion time relative to that of the explicit $\Hs$.}
\label{tab:ctov}
\small
\begin{tabular}{lrrrr}
\toprule
configuration & total & median & $\le$1\% & expansion \\
\midrule
$W=0$ & 60.5\% & 43.1\% & 6\% & 1.05 \\
$W=1000$ & 44.4\% & 13.5\% & 34\% & 1.12 \\
$W=10^4$ & 35.7\% & 0.0\% & 53\% & 1.17 \\
$W=10^5$ & 29.9\% & 0.0\% & 75\% & 1.23 \\
$1\cdot\mathit{out}+10$ & 49.3\% & 29.3\% & 16\% & 1.09 \\
$4\cdot\mathit{out}+10$ & 14.2\% & 1.4\% & 48\% & 1.19 \\
$16\cdot\mathit{out}+10$ & 8.9\% & 0.0\% & 61\% & 1.25 \\
$W=\infty$ & 0\% & 0\% & 100\% & 1.74 \\
\bottomrule
\end{tabular}
\end{table}

Figure~\ref{fig:ctov} and Table~\ref{tab:ctov} show the results on \ctov queries.
First, we note that the plain $\Hs$ is about 19 times larger than the differential one, but also 23 times faster, even if the latter runs on the faster plain graph representation. These are classic space-time tradeoffs of compressed representations, and serve different usage scenarios.

Things get more interesting in the lower space range: tunnelling with $W=0$ almost halves the space of compressed $\Hs$ and is {\em faster} to traverse (0.9 on Compact, 0.7 on Plain, and almost the same on the Ring), yielding a win-win situation. The reason is that the explicit
representation accesses each live neighbor individually, whereas a re-expanded node enumerates its
whole range of neighbors sequentially, which is cheaper when most of them are live. 

Using even less space, the relative budget $4 \cdot out+10$ is about as fast as the fixed budget $W=1000$, but the latter is much less effective:
in many large queries nodes have
wide fans of dead branches that die after one step. A tunnel cannot absorb them with a fixed
budget, but it can with one proportional to the many live edges the same classes produce. In addition, the relative budget is only slightly slower than the explicit differential $\Hs$: 1.19 times on the Ring and 1.01 times on Compact, while being 7.1 times smaller.

Against the extreme of using no space, the relative budget is faster than re-running the query from scratch by a factor of 1.45 on the Ring and 1.49 on Compact, at the price of storing about 5.2 KB of data per \ctov query, on average.

Larger budgets ($W=10^5$ and $16\cdot\mathit{out}+10$ in Table~\ref{tab:ctov}) reduce the
space further moderately (from {35.7\% to 29.9\%} and from {14.2\% to 8.9\%}), at a further cost
in time ({1.23 and 1.25}), so the gains diminish; we omit them in the
rest of the paper.

Across representations, Compact is smaller than the Ring (7.0 vs.\ 16.3 bytes per edge) and
always expands faster (e.g., 204 vs.\ 289 ms per query with $4\cdot\mathit{out}+10$), so it
dominates the Ring on these queries. Plain is about 7 times faster still (28 ms), yet 3 times larger than Compact. The choice
of representation is then another classic space-time tradeoff; tunneling works well on
all of them.

In short, for constant-to-variable queries 
tunnels multiply the savings of the differential encoding. Encoding $\Hs$ relative to the graph makes it about 19 times smaller than the plain $\Hs$, at the price of slower expansion. With $W=0$, tunnels almost halve it at no cost in time. With a relative budget, they shrink it by another order of magnitude (to 0.7\% of the plain $\Hs$ with $4\cdot\mathit{out}+10$, and less with a larger $c$) while expanding only 1.0--1.2 times slower than the explicit compressed $\Hs$, and about 1.5 times faster than re-running the query. These conclusions hold on the three graph representations, whose choice only shifts the overall speed.

\subsection{Constant-to-constant queries}
\label{sec:ctoc}

\begin{figure}[t]
\includegraphics[width=\linewidth]{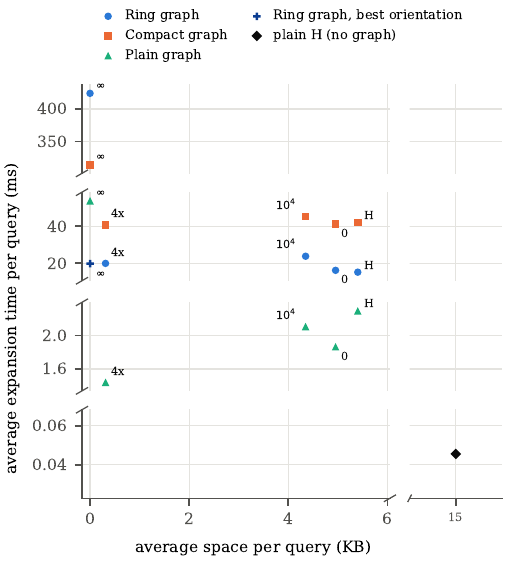}
\caption{Constant-to-constant queries, as in Figure~\ref{fig:ctov}.}
\Description{Scatter plot of average space per query against average expansion time per query for constant-to-constant queries, on the three graph representations, including the Ring with the best orientation and the plain solution subgraph.}
\label{fig:ctoc}
\end{figure}

Over \ctoc queries, $\R$ has on average 187 thousand nodes and $\Hs$ only 910 nodes (and about 4200 edges); the median query has an
$\Hs$ of 3 edges. Figure~\ref{fig:ctoc} shows a very different picture from \ctov queries. Re-running
from the source of the original query is prohibitive: 27.9 times the expansion of the explicit $\Hs$ on the Ring and 23.5 times on Plain, as most of the traversal is dead. The absolute budgets thus barely compress (80\% of the
explicit $\Hs$ with $W=10^4$). The relative budgets, instead, store only 5.8\% of it,
while expanding in 1.31 times the
time on the Ring and \emph{faster} than the explicit $\Hs$ on the other two representations: the
large $\Hs$s, which dominate the space, produce many live edges per unit of dead work.

The orientation of the traversal matters, however, much more than tunnels: going up a hierarchy each node has few parents, whereas
going down every node on the way to the target has many siblings whose subtrees are dead. We then explore an alternative solution:
we solve every pair from both endpoints. Starting from the one with the smaller product graph
reduces $\R$ from 187K nodes to 1,655 on average. 
One may think this is an artifact from having converted \ctov queries to \ctoc,
but we will see it again on the generated queries. 
With the best orientation, re-running the query takes
only 1.30 times the expansion of the explicit $\Hs$ (``best orientation'' in Figure~\ref{fig:ctoc}),
the same as tunnels with $4\cdot\mathit{out}+10$, which store 3.7\% of $\Hs$ (under 200 bytes).
For \ctoc queries, then, one bit per
pair telling the decoder where to start replaces the stored structure, and a decoder that
dovetails both orientations needs no space at all, at most doubling its time.

Unlike on \ctov queries, the Ring is faster than Compact here: it expands the explicit $\Hs$ in 15 milliseconds and $\Sset$ with $4\cdot\mathit{out}+10$ in 20, against 42 and 41 milliseconds on Compact.
Neither dominates, then: Compact takes less than half the space. Plain is again the fastest (1.45 milliseconds with $4\cdot\mathit{out}+10$) with the most space. The reason is that Compact
navigates one direction faster than the Ring and the other slower: per edge of the product graph,
its \ctov traversals take 5.4\,$\mu$s against 3.0 on the Ring, and its \vtoc traversals 1.4 against
2.3. For the same reason, it cannot properly exploit the idea of choosing the best direction for \ctoc queries.

In short, for constant-to-constant queries the orientation of the traversal matters more than tunnels. From the source of the query, most of the traversal is dead and re-running is tens of times slower than expanding 
\Hs, which tunnels with a relative budget avoid at a small cost in space. Starting from the right endpoint, instead, removes most of the dead work, so that re-running is about as fast as expanding 
\Hs, and one bit per pair telling the decoder where to start replaces the stored structure. Among the representations, the Ring is faster than Compact here, because these traversals mostly use the direction that Compact navigates more slowly.

\subsection{Variable-to-variable queries}
\label{sec:vtov}

\begin{figure}[t]
\includegraphics[width=\linewidth]{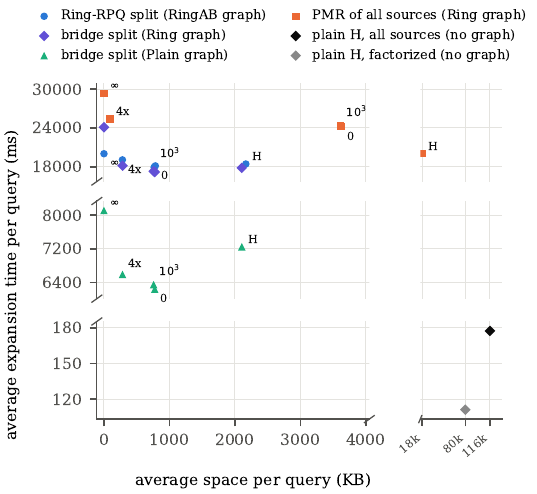}
\caption{Variable-to-variable queries: average space per query against
average expansion time per query of the factorized representation
with the two cuts, and of the PMR of all
sources. The bridge split on the Ring nearly coincides with the 
Ring-RPQ split on RingAB, so both are labeled once. The diamonds are the plain $\Hs$ of all sources and of the factorized representation. Labels as in
Figure~\ref{fig:ctov}, with $W=10^3$.}
\Description{Scatter plot of average space per query against average expansion time per query for variable-to-variable queries, comparing the factorized representation with two cuts, the PMR of all sources, and the plain solution subgraphs.}
\label{fig:vtov}
\end{figure}

Figure~\ref{fig:vtov} shows the 171 \vtov queries with a cut. The PMR of all sources and the
factorized representation have about the same nodes (7.7M per query), but without tunnels the PMR
of all sources takes 8.2 times more space (17.8 vs.\ 2.2 MB per query with differential
compression, and 116 vs.\ 80 MB in plain form), for the reason given in Section~\ref{sec:classes}:
its 6.9M candidate sources per query, against 74K seeds, are internal nodes. With tunnels the
comparison reverses: sources without dead successors need not be stored, and the PMR of all
sources takes 93 KB per query with $4\cdot\mathit{out}+10$, against 282 KB for the factorized
representation. Starting from far fewer nodes, however, the factorized representation is faster:
it expands in 19 instead of 25 seconds per query, re-runs in 20 instead of 29, and is built in 27
instead of 44. The PMR of all sources, in turn, also handles the 56 queries without a label to cut
at. The remaining 72 of the 299 queries match the empty path (as \texttt{a*} does), so every node is
a solution with itself; we exclude them.

Tunnels barely slow down the factorized representation: with $4\cdot\mathit{out}+10$ it takes 13\%
of its compressed form and expands in 1.03 times its time, and re-running it from scratch costs only
1.09 times, because its traversal starts from seeds that almost all lead to solutions. The two cuts
give almost the same representation (less than 3\% apart). The bridge split, however, is slower to
build (43 vs.\ 27 seconds per query) and to re-run (24 vs.\ 20): its seeds are fewer (58K vs.\ 74K per
query), but 16\% of them die, against 1\%, so its traversal explores 21\% more product nodes. The
Ring-RPQ split cuts at the intersection of two mandatory labels, chosen by selectivity, which filters
out the dead seeds. In exchange, the bridge split does not need RingAB, which takes 70\% more space.

Across representations, Plain expands the factorized representation about 3 times faster than the
Ring, whereas Compact is about 5 times slower (with $W=0$, on the 168 queries it completed before we
stopped the run). Unlike on \ctoc queries, this slowdown does not come mainly from the traversal
direction, but from operations that \vtov queries add: Compact lacks a range operation to count and
list the distinct sources or targets of a label, so choosing the bridge cut and regenerating the
seeds traverses all the edges of the label (35\% of the instructions in a profile on a sample of the
graph); the backward traversals of the sides add another 32\%.

In short, for \vtov queries the factorized representation is the fastest, and by far the smallest
without tunnels. With tunnels, both become small and their sizes are comparable: on these queries
the PMR of all sources is 3 times smaller, but on the generated ones it is somewhat larger
(see next); it is always slower to decode and build.

\subsection{Generated queries}
\label{sec:gen}

\begin{table}[t]
\caption{Generated queries on the Ring (RingAB for \vtov): space of $\Sset$ relative to the explicit
$\Hs$ with differential compression, and expansion time relative to that of the explicit $\Hs$. The
largest absolute budget is $W=10^4$ ($W=10^3$ for \vtov). The first row gives the average space and time per query.}
\label{tab:gen}
\small
\setlength{\tabcolsep}{3.5pt}
\begin{tabular}{lrrrrrr}
\toprule
 & \multicolumn{2}{c}{\ctov} & \multicolumn{2}{c}{\ctoc} & \multicolumn{2}{c}{\vtov} \\
\cmidrule(lr){2-3}\cmidrule(lr){4-5}\cmidrule(lr){6-7}
configuration & space & time & space & time & space & time \\
\midrule
explicit $\Hs$ & {88\,KB} & {459\,ms} & {1.1\,KB} & {1.2\,ms} & {2.1\,MB} & {20.4\,s} \\
$W=0$ & 49.5\% & 0.90 & 98.5\% & 0.94 & 21.8\% & 1.01 \\
largest $W$ & 49.1\% & 0.90 & 65.1\% & 1.64 & 21.1\% & 1.01 \\
$4\cdot\mathit{out}+10$ & 13.3\% & 0.99 & 19.1\% & 1.46 & 6.8\% & 1.05 \\
{$16\cdot\mathit{out}+10$} & 0.59\% & 1.04 & 16.7\% & 1.39 & 6.1\% & 1.07 \\
$W=\infty$ & 0\% & 1.11 & 0\% & 557 & 0\% & 1.10 \\
\bottomrule
\end{tabular}
\end{table}

Table~\ref{tab:gen} summarizes the generated queries. The comparisons exclude 5 \ctov queries,
all Kleene stars from the same hub constant, whose product graph exceeded our time limit on the Ring
(Plain built them; their $\Hs$s are 3.5 times larger than those of all the other queries together).
The results confirm those on the log, with some differences worth noting.

On \ctov queries, the plain $\Hs$ is 33 times larger than the compressed one (2.9 MB vs.\ 88 KB per
query). Most queries store almost nothing already with $W=0$ (median 0.1\% of $\Hs$), but a few
Kleene stars from hubs dominate the total: every live successor has a dead twin, the same neighbor in
another automaton state that has no continuation, so the dead work grows with the output and no
absolute budget absorbs it. The relative budget does, if $c$ is large enough. In the 19 queries of
shape \texttt{(a|b|c|d|e)*}, each live successor has up to four dead twins of cost 2, about 8 units
per edge of $\Hs$, so $4\cdot\mathit{out}+10$ stores 13.3\% of $\Hs$ (88\% of it in those
queries), whereas $16\cdot\mathit{out}+10$ stores 0.59\% (0.5 KB per query) and expands in 1.04
times the time of the explicit $\Hs$ (Appendix~\ref{app:shapes}). On the log, the larger $c$ also
helps, though less (Table~\ref{tab:ctov}).

On \ctoc queries, choosing the orientation works too: starting from the endpoint with the smaller
product graph reduces $\R$ from 548K to 3.1K nodes per pair, and re-running from there is 100 times
faster than from the other end. The explicit $\Hs$ of these pairs is so small, however,
that re-running still takes 5.2 times its expansion, so tunnels remain useful: with
with
$4\cdot\mathit{out}+10$ they store 11.7\% of $\Hs$ and expand in 1.46 times its time (9.3\% at 1.39 times with $16\cdot\mathit{out}+10$).

On \vtov queries, the factorized representation behaves as on the log: tunnels reduce it to 6.8\%
at 1.05 times its expansion time (6.1\% at 1.07 times with $16\cdot\mathit{out}+10$, so here a larger $c$ gains little), and on Plain the tunnelled form expands even 10\% faster than the
explicit one. The PMR of all sources takes 8.5 times more space than the factorized
representation without tunnels, and here also more with them (187 vs.\ 144 KB per query, although it
is smaller in 55\% of the queries), and it expands 1.34 times slower; it also handles 89 more
queries, which have no label to cut at.

Across representations, Compact again dominates the Ring on \ctov queries (7\% faster and 2.3 times
smaller), Plain is about 4.6 times faster than Compact, and Compact is again about 5 times slower
than the Ring on \vtov queries.

\section{Discussion and future work}
\label{sec:discussion}

We have shown that the solution subgraph $\Hs$ of an RPQ can be represented in a small
fraction of its size when the graph is available at decoding time. Encoding $\Hs$ relative to the
graph already reduces it about 20 times; tunnels then store only the nodes whose dead branches
are too costly to explore again, which with a budget proportional to the output reduces the space
by another order of magnitude (to about 14\% of the compressed $\Hs$ on \ctov queries) while
expanding in about the same time as the explicit $\Hs$. On \ctoc queries, re-running from the
source of the query can be hundreds of times slower than expanding, but starting from the right
endpoint makes it almost as fast, with nothing stored. For \vtov queries, factorizing the PMR of all sources at a
cut point makes it 8 times smaller and faster to build and expand, and tunnels reduce either to 0.1--0.3 MB per query. These results hold on the three graph
representations we tried, whose choice is an ordinary space-time tradeoff: a
plain representation is the fastest but uses over 20 bytes per edge; the best
compact representation (7 bytes per edge) for \ctov and \vtoc queries is Compact
\cite{NR24}; the Ring \cite{AHNRRS21} is faster than Compact on \ctoc and \vtov
queries but it uses 16 bytes per edge. 


Various optimizations are possible as future work, but probably the most
interesting one is that we build and decode $\Hs$ one product node at a time, whereas Ring-RPQ processes
all the states of a graph node at once, bit-parallelly. This could also be done for $\Hs$: a
forward pass marks in each graph node the states reachable from the root, and a backward pass
from the accepting nodes marks those that reach them; their intersection gives the nodes of
$\Hs$, and its edges are all the edges of the product graph between them, which the graph and the
automaton determine. Expanding a graph node with all its pending states at once could likewise make
the re-expansion of tunnels cheaper, which is precisely the cost that tunnels pay.
The same idea suggests a plain representation of $\Hs$ as a subgraph of $\G$ rather than of
the product graph: the graph edges used by $\Hs$, with their labels, plus the mask of live states of
each node. Edges and states are then recovered with the automaton, and the masks avoid following
an edge that is live only for another state.

\appendix

\section{Results per expression shape}
\label{app:shapes}

\begin{table}[t]
\caption{Generated \ctov queries on the Ring, for the 12 most frequent shapes: number of queries,
total edges of $\Hs$ (millions), explicit $\Hs$ (MB), space of $\Sset$ as a percentage of it, and
expansion time ($t$) relative to the explicit $\Hs$ with $4\cdot\mathit{out}+10$ (``4x'') and $W=\infty$.}
\label{tab:shapes}
\footnotesize
\setlength{\tabcolsep}{1.6pt}
\begin{tabular}{lrrrrrrrr}
\toprule
shape & $n$ & edges & $\Hs$ & $W{=}0$ & $W{=}10^4$ & 4x & $t_{\mathrm{4x}}$ & $t_\infty$ \\
\midrule
\texttt{a/b*} & 363 & 138.1 & 22.5 & 3.8\% & 3.3\% & 1.0\% & 0.85 & 0.90 \\
\texttt{a*} & 181 & 30.6 & 15.4 & 99.5\% & 98.8\% & 0.03\% & 1.40 & 1.41 \\
\texttt{a/b} & 132 & 6.1 & 5.5 & 1.4\% & 1.2\% & 0.12\% & 1.21 & 1.27 \\
\texttt{a|b} & 86 & 32.4 & 4.1 & 0\% & 0\% & 0\% & 0.86 & 0.84 \\
\texttt{\^{}a} & 25 & 8.7 & 1.1 & 0\% & 0\% & 0\% & 0.87 & 0.88 \\
\texttt{a+} & 25 & 19.6 & 9.8 & 100.0\% & 99.7\% & 0.01\% & 1.50 & 1.50 \\
\texttt{(a|b|c|d|e)*} & 19 & 20.5 & 10.3 & 99.9\% & 99.6\% & 99.8\% & 0.88 & 1.47 \\
\texttt{(a|b)*} & 16 & 3.2 & 1.6 & 99.7\% & 98.5\% & 0.09\% & 1.19 & 1.19 \\
\texttt{a/b/c*} & 13 & 2.1 & 3.5 & 0.38\% & 0.29\% & 0\% & 0.96 & 0.96 \\
\texttt{a/b+} & 13 & 4.3 & 2.1 & 12.8\% & 12.8\% & 12.8\% & 1.32 & 2.08 \\
\texttt{a|b|c|d} & 12 & 2.3 & 0.3 & 0\% & 0\% & 0\% & 0.88 & 0.89 \\
\texttt{a/b*/c*} & 9 & 0.5 & 0.1 & 3.5\% & 0\% & 0.41\% & 0.89 & 0.89 \\
\bottomrule
\end{tabular}
\end{table}

Stars from a constant (\texttt{a*}, \texttt{a+}, \texttt{(a|b|\dots)*}) store almost all of $\Hs$ with
any absolute budget, because every live successor has a dead twin in another automaton state, and
almost nothing with the relative budget, except \texttt{(a|b|c|d|e)*}: there each live
successor has up to four dead twins, one per other letter, each costing 2, so the dead work exceeds
$4\cdot\mathit{out}$. These 19 queries account for 88\% of the space that the relative budget
leaves on the generated \ctov queries. Shapes with a fixed first label, such as \texttt{a/b*}
or \texttt{a/b}, store little already with $W=0$. Alternatives and single labels (\texttt{a|b},
\texttt{\^{}a}) have no dead work at all. The 12.8\% of \texttt{a/b+} comes from a single query
from a hub, whose root is stored with every budget, with a list of 387K successors (265 KB). Its
dead branches add up to 9M product nodes, three times the size of $\Hs$, so the dead work exceeds
even the relative budget; raising the cap on the exploration of a dead branch up to $10^7$ does not
change this. Storing the root makes the expansion take 23.4 instead of 36.0 seconds (17.9 for the
explicit $\Hs$).

\bibliographystyle{ACM-Reference-Format}
\bibliography{tunnels}

\balance

\end{document}